\documentclass[twocolumn,amsmath,amssymb,floatfix,groupedaddress,prl]{revtex4}
\usepackage{graphicx}
\usepackage{color}
\usepackage{ulem}
\usepackage{hyperref}

\begin{document}

\title{Surface-plasmon mode hybridization in sub-wavelength microdisk lasers}

\author{R. Perahia}
\email{rperahia@caltech.edu}
\author{T. P. Mayer Alegre}
\author{A. Safavi-Naeini}
\author{O. Painter}
\affiliation{Thomas J. Watson, Sr., Laboratory of Applied Physics, California Institute of Technology, Pasadena, CA 91125}
\date{\today}

\begin{abstract} 
Hybridization of surface-plasmon and dielectric waveguide whispering-gallery modes are demonstrated in a semiconductor microdisk laser cavity of sub-wavelength proportions.  A metal layer is deposited on top of the semiconductor microdisk, the radius of which is systematically varied to enable mode hybridization between surface-plasmon and dielectric modes. The anti-crossing behavior of the two cavity mode types is experimentally observed via photoluminescence spectroscopy and optically pumped lasing action at a wavelength of $\lambda \sim 1.3$~$\mu$m is achieved at room temperature.
\end{abstract}

\maketitle

In wavelength-scale lasers, the very small number of optical modes and small volume of gain material allows one to probe the subtle and often interesting properties of lasing action \cite{ref:Yamamoto2}.  Semiconductor microdisk lasers, in particularly, have been actively studied due to their simple geometry and amenability to planar chip-scale integration with microelectronics\cite{ref:Levi,ref:Baba6,ref:Painter3}.  More recently there has been great interest in using surface-plasmon (SP) modes at a semiconductor-metal interface for guiding as well as high intensity and sub-wavelength optical confinement \cite{ref:Ebbesen}. There has been significant work on the incorporation of SP waveguides that also act as electrical contacts in mid-infrared quantum cascade lasers \cite{ref:Sirtori2}, in increasing SP propagation lengths using SP-dielectric waveguide mode hybridization \cite{ref:Min_BK}, as well as in creating ultra-small laser cavities \cite{ref:Hill_MT,ref:Hill_MT2}.

\begin{figure}[t]
\begin{center}
\includegraphics[width=\columnwidth]{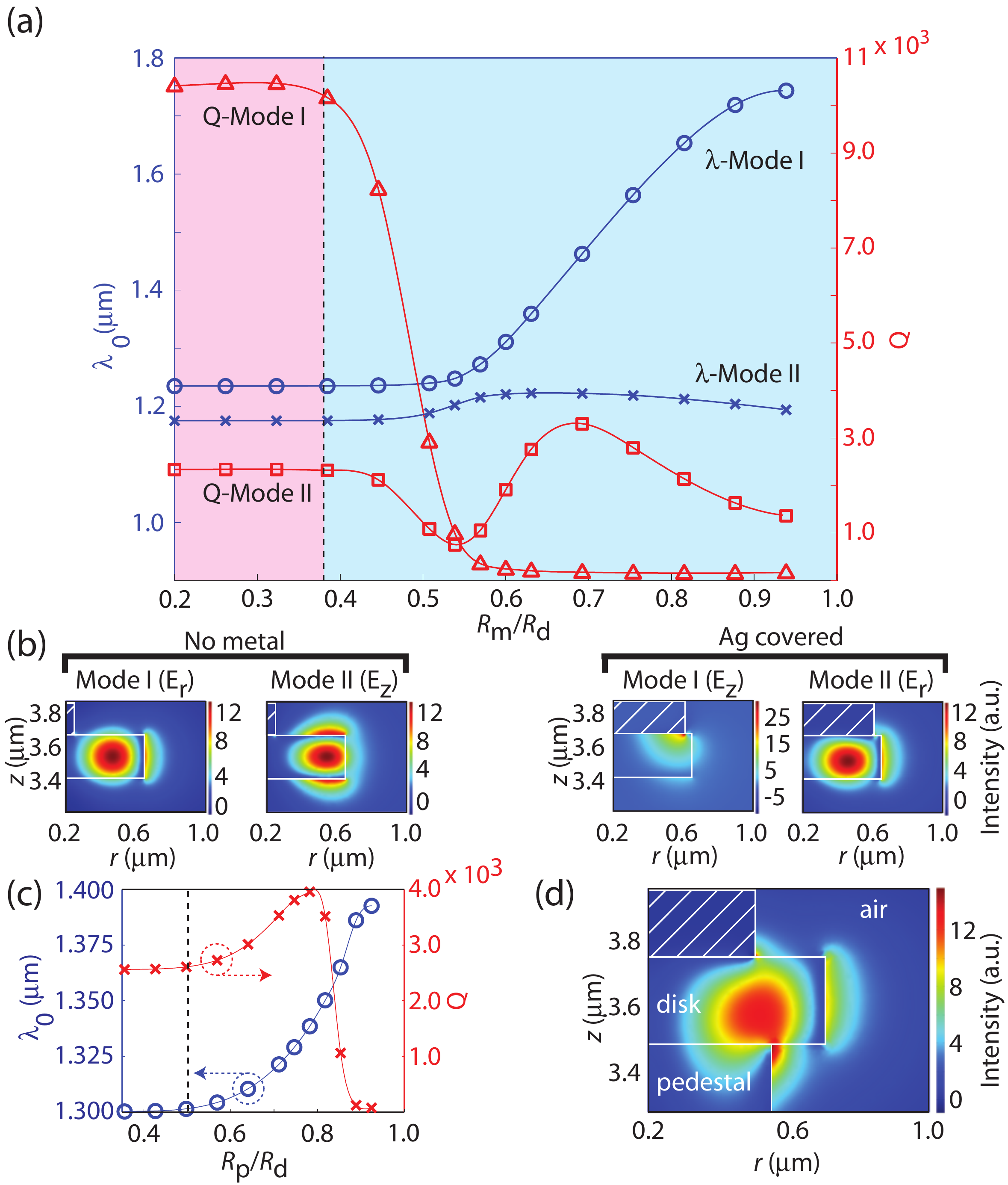}
\caption{A $250$~nm thick semiconductor microdisk with radius $R_d = 0.65$~$\mu$m and a top silver contact is simulated with varying silver coverage. (a) Wavelength ($\lambda_0$) and quality factor ($Q$) for two anti-crossing modes (mode I and mode II) are plotted as a function of silver metal radial fraction ($R_m/R_d$). (b) Azimuthal slices of dominant electric field components of the two extreme cases: with and without silver. Metal is denoted by white hatch marks. (c) Wavelength and $Q$ of mode II as a function of increasing InP pedestal radius ($R_p$) with $R_{m}/R_{d} = 0.7$. (d) Mode profile of $E_{r}$ field component of mode II with maximum $Q$ ($R_{p}/R_{d} = 0.8$.)}
\label{fig:recede_metal_simulation}
\end{center}
\end{figure}

In miniaturizing semiconductor lasers to the nano-scale one encounters several design challenges that must be addressed, such as thermal management\cite{ref:Shih_MH,ref:Hwang_JK}, proximity of metal contacts to the optical cavity,  surface states\cite{ref:Tai_K}, and demanding tolerance levels in fabrication. In this Letter we investigate the purposeful integration of a metal contact into a sub-wavelength whispering-gallery microdisk laser.  We show that whispering-gallery SP and dielectric modes hybridize into low loss modes. We predict and map out this hybridization using finite-element-method (FEM) simulations, and experimentally measure the properties of fabricated microdisk laser cavities with varying levels of mode hybridization.
\begin{figure}[h]
\begin{center}
\includegraphics[width=\columnwidth]{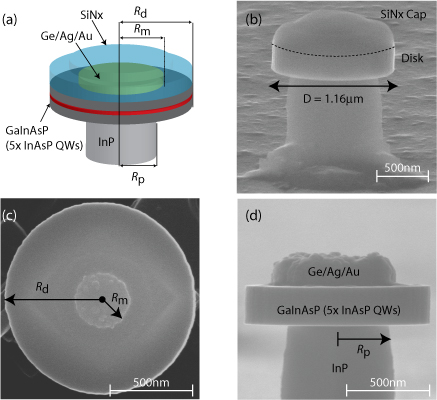}
\caption{(a) Schematic of fabricated and simulated microdisks.  (b-d) Scanning electron microscope (SEM) images of a diameter $D = 1.16$~$\mu$m fabricated microdisk with a Ge/Ag/Au $10/80/10$~nm contact buried under a $150$nm conformal layer of SiN$_x$. Dashed line delineates GaInAsP--SiN$_x$ boundary. (c) Top view and (d) cross-sectional view after SiN$_x$ has been removed.}  
\label{fig:geometry_a}
\end{center}
\end{figure}
Simulation of the hybrid laser cavities is performed using fully-three-dimensional FEM simulations with azimuthal symmetry~\cite{ref:Borselli3}. A $250$~nm thick semiconductor disk with index $n_{disk} = 3.4$ and radius $R_{d} = 0.65$~$\mu$m is simulated with a centered metal contact of varying radius ($R_m$).  A schematic of the microdisk device is shown in Fig.~\ref{fig:geometry_a}(a).  Silver with an imaginary refractive index of $n_{Ag} = 0.11 -i9.5$ at $\lambda = 1.3$~$\mu$m is chosen for the metal layers due to its low optical loss~\cite{ref:Christy_RW}.  For this disk size in the $1300$~nm wavelength band there occurs a near degeneracy of the transverse-electric-like (TE-like) whispering-gallery mode (WGM) with dominant electric field polarization in the plane of the disk and the transverse-magnetic-like (TM-like) mode with dominant electric field normal to the disk plane.  A plot of the wavelength and $Q$-factor of these two resonances is shown in Fig. \ref{fig:recede_metal_simulation}(a) as a function of metal radius fraction ($R_{m}/R_{d}$).  The resonances exhibit a clear anti-crossing behavior, with the modes hybridizing and picking up significant SP character with increased metal coverage.  At one extreme, where $R_{m}/R_{d}$ is very small, the upper wavelength branch (mode I) is of TE-like WGM character and the lower wavelength branch (mode II) is of TM-like WGM character.  At the other extreme, $R_{m}/R_{d} \sim 1$, the upper branch has taken on a SP mode character whereas the lower branch is now TE-like.  Dominant electric field components for both cases are plotted in Fig. \ref{fig:recede_metal_simulation}(b).  

Of particular interest would be the lower wavelength branch (mode II) of the SP-dielectric hybrid modes, as this mode shows significant robustness in its optical $Q$-factor for large metal coverage.  Further analysis of this regime is performed by studying the effects of an InP pedestal ($n = 3.2$) used to support the microdisk.  Figure~\ref{fig:recede_metal_simulation}(c) shows a plot of cavity $Q$ of mode II versus the pedestal radius ($R_{p}$) with $R_{d} = 0.7\mu$m so the resonant wavelength is $\lambda \approx 1.3$~$\mu$m.  In this plot the metal-to-disk ratio was set at the optimal value of $R_{m}/R_{d} = 0.7$ from Figure~\ref{fig:recede_metal_simulation}(a).  A peak value of $Q = 4000$ is found for a fractional pedestal radius of $R_{p}/R_{d} = 0.8$ (corresponding mode field plot shown in Fig \ref{fig:recede_metal_simulation}(d)).  From these simulations it is clear that a microdisk cavity structure with significant metal coverage and very little undercut can be designed to have a cavity $Q$-factor sufficient for lasing action.  Such a structure would facilitate fabrication as well as good thermal heat sinking.  

\begin{figure}[t]
\begin{center}
\includegraphics[width=\columnwidth]{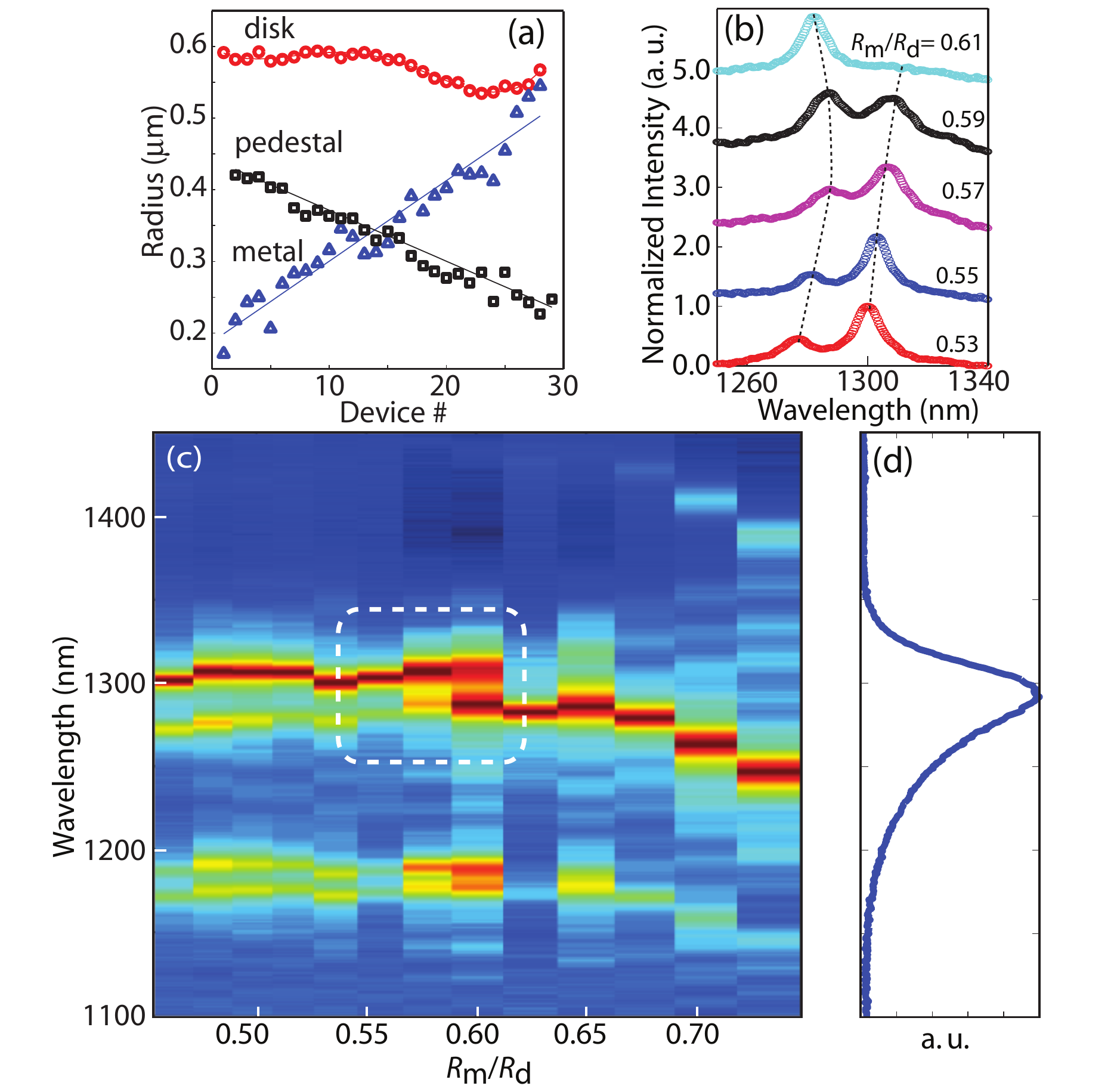}
\caption{(a) Measured disk, metal, and pedestal radii of fabricated device array.  (b) Zoom-in of anti-crossing region as indicated by a dashed white box in panel (c). Dotted lines are guides to the eye.  (c) Fiber-taper-collected normalized spectra on a log-scale as a function of $R_{m}/R_{d}$ ($P_{p}=1$~mW, $\delta T = 500$~ns, $T=1$~$\mu$s). (d) Free-space-collected PL of unpatterned laser material ($P_{p} = 282$~$\mu$W, continuous-wave).}
\label{fig:anti_crossing}
\end{center}
\end{figure}

In order to test the predictions of the FEM-modeling, microdisks with nominal diameter $D = 1.2$~$\mu$m were fabricated from $252$~nm thick membranes consisting on five InAsP/GaInAsP compressively strained quantum wells, with peak spontaneous emission at $\lambda \approx 1285$~nm \cite{ref:Hwang2}. First, the metal layer (Ge/Ag/Au=$10/80/10$~nm) was deposited and patterned into round contacts of systematically varying diameter by electron beam lithography (EBL), electron beam evaporation, and liftoff.  A hard SiN$_x$ mask was then deposited to protect the patterned metal layer, and the the outer disk shape was patterned by a second, aligned EBL step.  An inductively-coupled reactive-ion etch was used to transfer the disk pattern through the hard mask and the $252$~nm thick semiconductor layer.  The disks were then undercut, and the pedestal formed, using HCl:H${_2}$O solution \cite{ref:Srinivasan5}.  A scanning electron microscope (SEM) micrograph of a final device including remaining SiN$_x$ cap is shown in Fig. \ref{fig:geometry_a}(b).  After the device testing described below, the SiN$_x$ cap layer was removed allowing for SEM imaging (Fig. \ref{fig:geometry_a}(c-d)) and measurement of the disk, metal, and pedestal radii (Fig.~\ref{fig:anti_crossing}(a)).  A systematic variation of $25$ to $100\%$ metal coverage was achieved in disks with average diameter $D = 1.2$~$\mu$m.


Initial resonance mode spectroscopy was performed by free space optical pumping and collection of the photoluminescence (PL) through an optical fiber taper nanoprobe~\cite{ref:Michael_CP1}.  The fiber taper provides excellent collection efficiency of the WGM emission of the microdisk, substantially improving the sensitivity of the measurement.  Using a pulsed external cavity diode laser at $\lambda = 830$~nm, the disks were pumped with pulses of peak power $P_{p} \approx 1$~mW, pulse width $\delta T = 500$~ns, and pulse period $T = 1$~$\mu$s.  The fiber-collected spectra are plotted in Fig.~\ref{fig:anti_crossing}(c) for devices with varying $R_{m}/R_{d}$. Two pairs of modes can be seen at $\lambda\sim 1.2$~$\mu$m and $\lambda \sim 1.3$~$\mu$m.  A normalized PL spectrum taken with continuous wave pumping from the unprocessed semiconductor material ($P_{p} = 282$~$\mu$W) is shown in Fig. \ref{fig:anti_crossing}(d) for reference.  Despite the disappearance of the upper wavelength branch mode in the spectra of Fig~\ref{fig:recede_metal_simulation}(c) for large metal coverage, a result expected due to the very low-$Q$ of the SP mode, the zoom-in shown in Fig.~\ref{fig:anti_crossing}(b) for the longer-wavelength mode pair shows clear anti-crossing behavior indicative of mode hybridization of the SP and dielectric WGMs.

\begin{figure}[t]
\begin{center}
\includegraphics[width=\columnwidth]{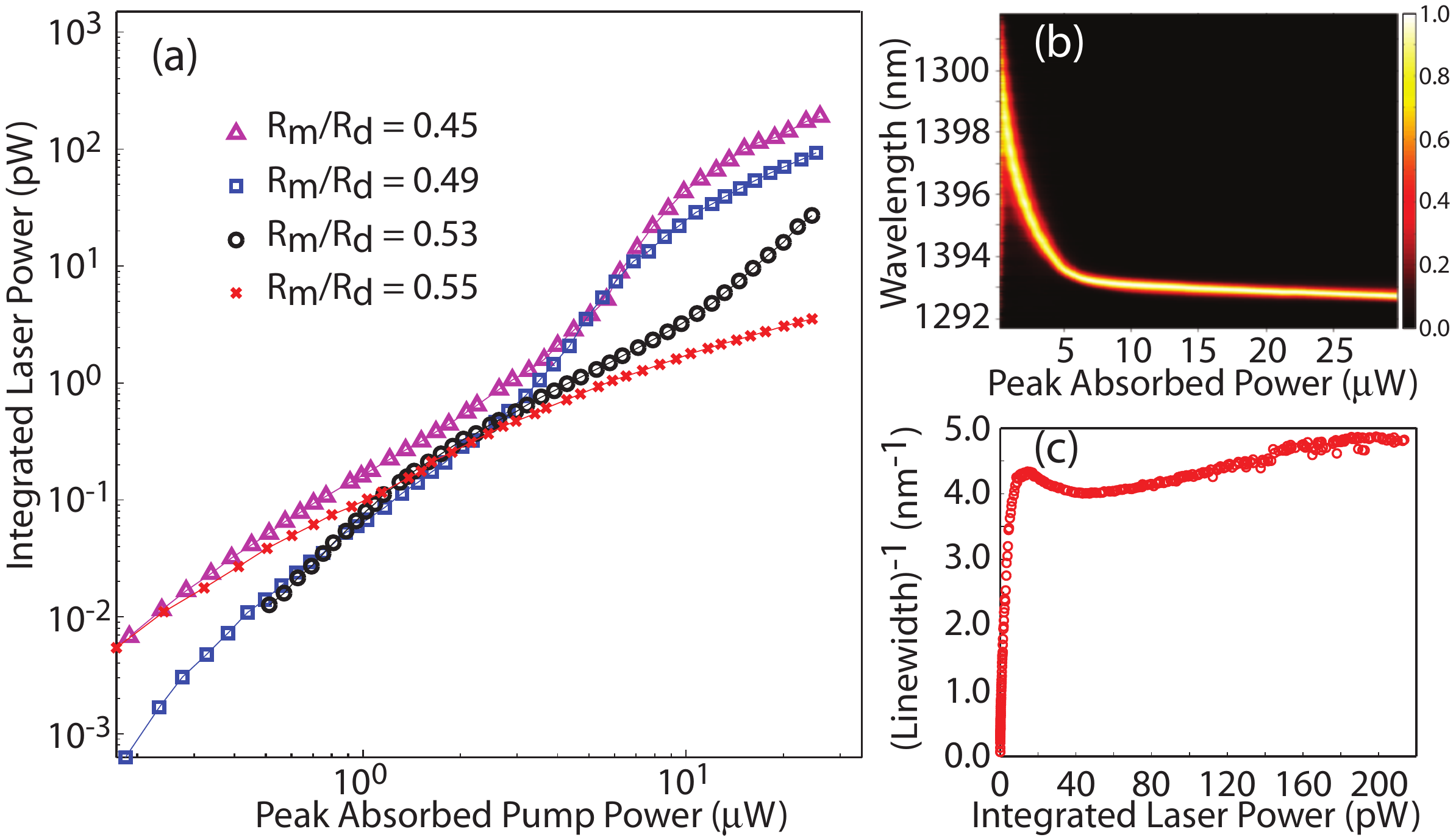}
\caption{(a) L-L curves of several devices of increasing $R_{m}/R_{d}$. (b) Normalized spectra versus pump power and (c) inverse linewidth versus output laser power for a device with $R_{m}/R_{d} = 0.42$.}
\label{fig:laser}
\end{center}
\end{figure}

To study laser action in these hybridized cavities the fiber taper is removed to eliminate external cavity loading effects, and vertically-scattered light emission from the microdisks is collected via a high numerical aperature lens instead.  The microdisks are also pumped with low-duty-cycle pulses ($\delta T = 20$~ns, $T = 4$~$\mu$s) in order to reduce thermal effects. Threshold curves for a series of microdisks near the region of strong SP and dielectric mode hybridization are shown in Fig.~\ref{fig:laser}(a).  The peak absorbed pump power is estimated based on a pump spot size of diameter $D = 2$~$\mu$m, absorption efficiency of $\eta = 10\%$, and accounting for the pump duty cycle. Disks with $R_{m}/R_{d}<0.53$ exhibit a clear ``S'' shaped logarithmic light-in versus light-out curve, indicating lasing action.  In each of these cases, the lasing mode is from the longer wavelength branch of the hybridized modes.  Typical of the laser behaviour for these devices is the laser with $R_{m}/R_{d} = 0.42$, which has an estimated threshold peak absorbed pump power of only $P = 5$~$\mu$W.  Spectra as a function of peak absorbed pump power for this laser are plotted in Fig. \ref{fig:laser}(b). A strong blue-shift of the laser wavelength with increased pumping is seen, attributable to free-carrier dispersion. The inverse linewidth as a function of integrated output power is also plotted in Fig. \ref{fig:laser}(c), and shows linewidth narrowing typical of a semiconductor laser with large coupling between carrier density (gain) and refractive index (cavity frequency)~\cite{ref:Agrawal_G2,ref:Bjork4}.  In the middle of the anti-crossing region, for microdisks with $0.55 \le R_{m}/R_{d} \le 0.65$, no lasing was observed.  This is likely due to the increased optical loss in this region predicted by simulation (see Fig.~\ref{fig:recede_metal_simulation}(a)).  For microdisks with $R_{m}/R_{d} > 0.65$ lasing action was also not observed, even for the higher-$Q$ shorter wavelength branch of modes.  Although simulations indicate that the optical cavity $Q$-factor for this mode should recover (and increase for an optimal pedestal size), it is likely that the corresponding shift in the mode to shorter wavelengths, and away from the gain peak, precludes lasing action.  

Beyond the initial demonstration of SP-dielectric mode hybridization and lasing in sub-wavelength partially metal coated microdisk cavities presented here, future efforts will focus on achieving electrically injected lasing action using the surface-plasmon metal layer as a top metal contact and thermal heat sink.  It is anticipated that further device engineering involving the top metal contact layer should allow room temperature, continuous-wave lasing action in such sub-wavelength laser cavities.  An improved optical quality factor may also be achieved in the shorter-wavelength mode branch by engineering the epitaxy thickness.  Ultimately, such improved structures should allow for the study of fundamental as well as practical issues associated with the scaling laws of deep-sub-wavelength mode volume semiconductor lasers, where the small-scale system size of the laser result in significant photon number and carrier number fluctuations~\cite{ref:Rice2}.   

This work was supported by the DARPA NACHOS program.  The authors would like to thank Kartik Srinivasan for helpful disucssion regarding the device processing and Jianxin Chen for growth of the laser material.


\end{document}